\documentclass[aps,pra,reprint,amsmath,amssymb]{revtex4-2}
\usepackage{graphicx}
\usepackage{physics}
\usepackage{xcolor}
\usepackage{tikz}
\usepackage{subfigure} 
\usepackage{hyperref}

\usetikzlibrary{shapes, positioning,decorations.pathmorphing, decorations.pathreplacing, arrows.meta}

\usepackage{quantikz}
\usepackage[left]{lineno}
\definecolor{bt}{RGB}{128, 0, 255}
\definecolor{erb}{RGB}{0, 56, 168} 

\definecolor{mma1}{RGB}{83,118,172}
\definecolor{mma2}{RGB}{220,145,33}
\definecolor{mma3}{RGB}{132,167,44}
\definecolor{mma4}{RGB}{232,88,47}
\definecolor{mma5}{RGB}{124,109,170}
\definecolor{mma6}{RGB}{189,99,25}

\usepackage{titlesec}

\begin{document}

\title{
Statistical Control of Relaxation and Synchronization in Open Anyonic Systems}

\author{Eric R Bittner}
\address{Department of Physics, University of Houston}
\author{Bhavay Tyagi}
\address{Department of Physics, University of Houston}

\date{\today}

\begin{abstract}

Quantum statistics dictate how particles exchange and correlate—but in two-dimensional systems, these rules extend beyond bosons and fermions to anyons, quasiparticles with continuously tunable exchange phases. Here, we develop a Lindblad framework for anyonic oscillators and show that fractional statistics enable statistical control of decoherence in open quantum systems. By varying the anyonic phase and environmental correlations, we demonstrate tunable mode protection, identify exceptional points in the dissipative spectrum, and reveal temperature-dependent coherence bifurcations. Using coherent multidimensional spectroscopy as a probe, we show that statistical phases leave distinct fingerprints in the third-order response, opening new routes to detect and manipulate topological excitations. These results establish the exchange phase as a functional control parameter for engineering dissipation-resilient quantum states.
\end{abstract}

\maketitle

\section{Introduction}

Anyons are quasiparticles in two-dimensional systems that obey fractional
exchange statistics, interpolating continuously between bosonic and fermionic
limits. These nontrivial exchange phases introduce unique dynamical signatures,
particularly in open quantum systems where environmental coupling influences
coherence and relaxation.
Fractional quantization permits particles in two dimensions to exhibit
statistics beyond the symmetric (bosonic) and antisymmetric (fermionic) wave-
function classes allowed in three dimensions. The theoretical groundwork was
established by Leinaas and Myrheim, who analyzed quantum mechanics in multiply
connected configuration spaces~\cite{Leinaas1977}. Wilczek later introduced a
tangible realization in which charged particles bound to magnetic flux tubes
acquire a phase upon exchange, providing the first physical model for anyons~\cite{Wilczek1982,Wilczek1982b}.

Goldin, Menikoff, and Sharp formalized this picture algebraically by classifying
inequivalent representations of the local current algebra in non-simply connected
spaces~\cite{Goldin1981}. These developments laid the foundation for understanding
topological phases of matter, particularly in the context of the fractional quantum
Hall effect (FQHE), where quasiparticles exhibit fractional charge and anyonic
statistics~\cite{Arovas1984}.

Experimental signatures of anyonic behavior have been observed in FQHE systems
through tunneling conductance, shot noise, and interferometric measurements.
Fractional charge values such as $e/3$ and exchange-phase-dependent interference
fringes provide compelling evidence for anyon statistics~\cite{Willett2009,Willett2013,Nakamura2020}.
Other platforms such as Kitaev’s toric code and topological superconductors also
support Abelian and non-Abelian anyons, with recent focus on Majorana zero modes
as a route to fault-tolerant topological quantum computation~\cite{Kitaev2003,Sarma2015,Nayak2008}.

More recently, attention has shifted toward understanding how anyonic systems
behave in open-system settings, where decoherence and dissipation play a central
role. Recent studies have shown that dissipation can drive topological phase
transitions, as in spin liquid systems undergoing anyon condensation~\cite{Yan2021}.
In Kitaev-type chains with boundary dissipation, anyon excitations display statistical-
phase-dependent asymmetries in relaxation and propagation~\cite{Wang2022}. These
findings motivate a detailed investigation into how exchange statistics and environmental
correlations jointly shape the coherence properties of open quantum systems.

We recently investigated noise-induced synchronization in open quantum systems
consisting of two coupled quantum oscillators interacting with a common, correlated
dissipative environment~\cite{Bittner2024b,Tyagi:2024aa,Bittner:2025aa} where
we assume that the environmental noise channels \( \mathbf{E}_1(t) \) and \( \mathbf{E}_2(t) \) evolve according to a coupled Ornstein–Uhlenbeck (OU) process written in differential form:
\begin{equation}
    \dd{\mathbf{E}}(t) = -\gamma \mathbf{E}(t)\dd{t} + \mathbf{B} \cdot \dd{\mathbf{W}}(t), \nonumber
\end{equation}
where \( \mathbf{E}(t) = \begin{pmatrix} E_1(t) \\ E_2(t) \end{pmatrix} \), and \( \dd{\mathbf{W}}(t) \) is a vector-valued Wiener process with correlation structure:
\begin{equation}
    \mathbb{E}\left[ \dd{\mathbf{W}}(t)\dd{\mathbf{W}}^\intercal(t') \right] = \boldsymbol{\Xi} \delta(t - t') \dd{t},
    \quad
    \boldsymbol{\Xi} =
    \begin{pmatrix}
    1 & \xi \\
    \xi & 1
    \end{pmatrix}. \nonumber
\end{equation}
Here, the parameter \( \xi \in [-1, 1] \) controls the degree of instantaneous correlation between the two noise channels. The matrix \( \mathbf{B} \) determines the noise amplitude and, together with \( \boldsymbol{\Xi} \), sets the stationary variance of the process. In the limit of perfect environmental correlation ($\xi = \pm 1$), the oscillators
exhibit long-lived phase synchronization, particularly when they are spectrally
near-resonant. The presence of nonzero \emph{quantum discord} between subsystems
indicated that correlated environmental noise can generate persistent quantum
coherences even in the absence of direct coupling. We further showed that two spins
can synchronize their relative phases through shared stochastic driving, akin to
the escapement mechanism that regulates motion in classical pendulum clocks.

In this study, we extend these concepts to anyonic systems and examine whether
synchronization and phase-locking behavior persist when particles obey fractional
statistics. We begin by developing an algebraic and dissipative framework for
anyon oscillators subject to correlated noise. By generalizing the bosonic and
fermionic algebras, we derive a Lindblad description that incorporates the anyonic
exchange phase $\theta$ as a continuous control parameter. We analyze the thermal
and dynamical properties of such systems, demonstrating that fractional statistics
induce a continuous deformation in the dissipative structure. This deformation
leads to tunable coherence bifurcations, exceptional points, and phase-sensitive
mode protection. We conclude by showing how manipulation of the statistical phase
and environmental correlations can be used to engineer dissipation-resilient
quantum states, a key requirement for robust quantum information processing.

\section{Theoretical Development}

We begin by developing a Lindblad framework for anyonic oscillators with fractional-statistics,
where the algebra of creation and annihilation operators is deformed by an
exchange phase $\theta$ that interpolates continuously between bosonic and
fermionic limits. For a single anyon oscillator, the operators $a^\dagger$ and $a$
satisfy the $q$-deformed commutation relation:
\[
a a^\dagger - e^{i\theta} a^\dagger a = 1.
\]
This deformation encodes the statistical phase acquired under exchange, with
$\theta = 0$ and $\theta = \pi$ recovering the bosonic and fermionic cases,
respectively. In a system of two anyon oscillators, the braided exchange relation
becomes $a_1 a_2 = e^{i\theta} a_2 a_1$, ensuring that the many-body wavefunction
acquires a phase $e^{i\theta}$ when particles are exchanged.

Abelian anyons, whose braiding statistics depend only on the number (but not order)
of exchanges, can be treated algebraically in a relatively straightforward manner.
Non-Abelian anyons, on the other hand, possess far richer and more complex exchange statistics. When two such particles are braided, the quantum state of the system evolves through a unitary transformation within a degenerate Hilbert space, meaning the outcome of braiding depends on the specific sequence of operations. This non-commutative behavior makes them especially promising for topological quantum computing, where information can be encoded and manipulated in a fault-tolerant way by controlling the braiding paths. 
Here, we focus on the open-system dynamics of Abelian anyons, reserving the
non-Abelian case for future work.

\subsection{Anyon master equation}
We first consider the dynamics of a single anyon oscillator coupled to a dissipative bath using the adjoint Lindblad formalism. The system is governed by the Hamiltonian \( H = \hbar\omega a^\dagger a \), where \( a \) and \( a^\dagger \) are the anyon annihilation and creation operators, respectively. Dissipative coupling is introduced via two Lindblad jump operators: \( L_{\text{em}} = \sqrt{\gamma(n_\theta + 1)}\, a \) for spontaneous and stimulated emission, and \( L_{\text{abs}} = \sqrt{\gamma n_\theta}\, a^\dagger \) for absorption. The thermal occupation number
\begin{align*}
n_{\theta} = \frac{1}{e^{\beta \omega} - e^{i \theta}}   
\end{align*}
may be complex due to the anyonic exchange phase \( \theta \).

In Appendix \ref{Appendix:A} we show how to derive the Lindblad equations of motion for anyonic operators $a$ and $a^\dagger$, which describe the coherences within the anyonic oscillator. In this, we adopt the approximate form $[H, a] \approx -\hbar\omega a$ to represent coherent evolution under a diagonal Hamiltonian $H = \hbar\omega \hat{N}$, where $\hat{N}=a^\dagger a$. While this is exact in the bosonic limit, it remains a valid approximation when the deformation from bosonic statistics is modest i.e., the thermal occupation is low (i.e., $n_\theta \ll 1$). A full derivation using the deformed algebra $[a, a^\dagger] = \Phi(\hat{N})$ shows that the Heisenberg equation acquires a number-dependent prefactor $\Phi(\hat{N})$, which introduces nonlinear corrections to the unitary dynamics. These corrections are neglected here to focus on the dissipative effects of fractional statistics. A detailed justification of this approximation is given in Section 3 of the Appendix \ref{Appendix:A}, where we derive the exact Heisenberg equation of motion and demonstrate the conditions under which the approximation holds.

\begin{figure}
    \centering
\includegraphics[width=0.95\linewidth]{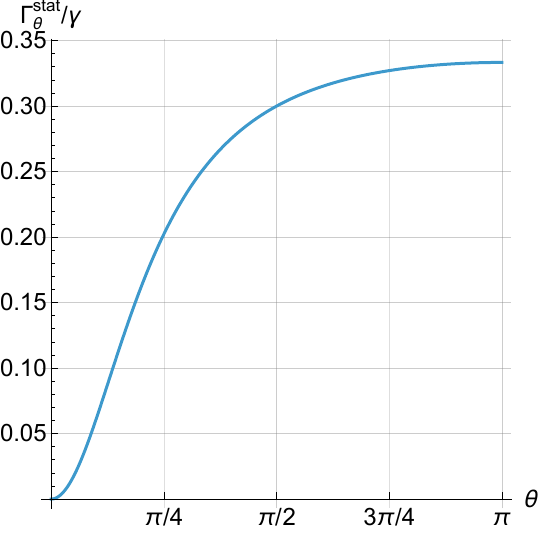}
    \caption{
Statistical contribution to the coherence relaxation rate as a function of the anyonic phase angle $\theta$. The rate vanishes in the bosonic limit ($\theta = 0$) and reaches a maximum in the fermionic limit ($\theta = \pi$). Unless stated otherwise, calculations use parameters $J = 0.2\omega$, $\gamma/\omega = 0.1$, and $\beta/\hbar\omega = 1$.
}
       \label{fig:1}
\end{figure}

Taking the expectation value over a thermal ensemble and including both thermal noise and statistical interference, the full coherence equation becomes:
\begin{equation}
\frac{d}{dt} \langle a \rangle = \left( -i \omega - \Gamma_\text{full} \right) \langle a \rangle,
\end{equation}
where the total phase relaxation rate is
\begin{equation}
\Gamma_\text{full} = \frac{\gamma}{2} \left[ 2 n_\theta + 1 + \left( 1 - \text{Re} \langle e^{i N \theta} \rangle \right) \right],
\end{equation}
where \( \langle e^{i N \theta} \rangle \) is the thermal expectation value of the anyonic phase factor given by 

\begin{equation}
\langle e^{i N \theta} \rangle = \frac{1 - e^{-\beta \omega}}{1 - e^{i \theta} e^{-\beta \omega}},
\end{equation}
which determines the phase-dependence of the steady-state dynamics. As \( \theta \to 0 \), this expression reduces to 1, recovering the bosonic limit. As \( \theta \to \pi \), it corresponds to the fermionic relaxation structure.
As result, the phase-dependent relaxation rate due purely to statistical effects is  given by
\begin{equation}
\Gamma_\theta^{\text{(stat)}} = \frac{\gamma}{2} \left( 1 - \text{Re} \langle e^{i N \theta} \rangle \right),
\end{equation}
where
\begin{equation}
\langle e^{i N \theta} \rangle = \frac{1 - z}{1 - z e^{i \theta}}, \quad z = e^{-\beta\hbar \omega}.
\end{equation}
Taking the real part yields
\begin{equation}
\text{Re} \langle e^{i N \theta} \rangle = \frac{(1 - z)(1 - z \cos \theta)}{1 - 2z \cos \theta + z^2},
\end{equation}
and hence the explicit form of the relaxation rate is
\begin{equation}
\Gamma_\theta^{\text{(stat)}} = \frac{\gamma}{2} \left[ 1 - \frac{(1 - z)(1 - z \cos \theta)}{1 - 2z \cos \theta + z^2} \right].
\end{equation}
This result reflects only the suppression of coherence due to exchange statistics. Notably, it vanishes for \( \theta = 0 \), implying no phase relaxation in the bosonic limit within this simplified statistical model. However, this does not imply that bosonic systems are immune to decoherence. In reality, thermal noise contributes to the decay of coherence even for bosons.
The full bosonic coherence decay rate, derived from the Lindblad equation with thermal bath coupling, is given by
\begin{equation}
\Gamma_\text{boson}^{\text{(thermal)}} = \frac{1}{2} \gamma (2 n_\theta + 1),\label{eq:8}
\end{equation}
which remains finite at \( \theta = 0 \)
(c.f. Fig.~\ref{fig:1}).

For bosonic systems ($\theta = 0$), 
the statistical contribution vanishes, leaving only the thermal component. 
As the anyon phase increases towards 
the fermonic regime, the stastistical 
contribution increase monotonically
until its maximum value at $\theta = \pi$ for a purely fermonic system. 
Consequently, moving from 
a purely fermionic system towards a fractionally quantized system results
in a decrease in relaxation rate of 
the oscillator at a given temperature. 

\begin{figure*}[t]
\subfigure[]{
\includegraphics[width=0.45\linewidth]{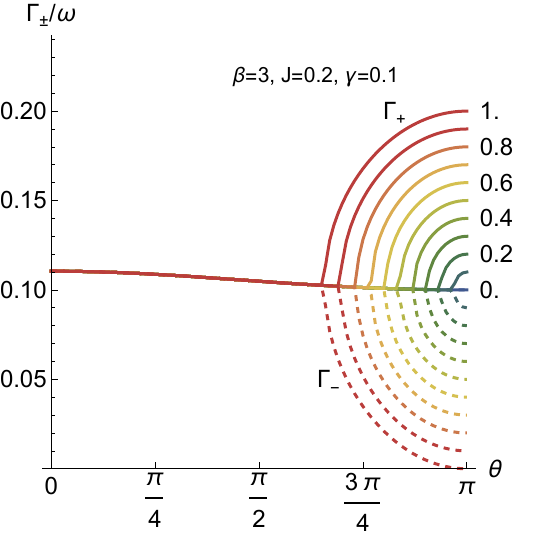}}
\subfigure[]{\includegraphics[width=0.45\linewidth]{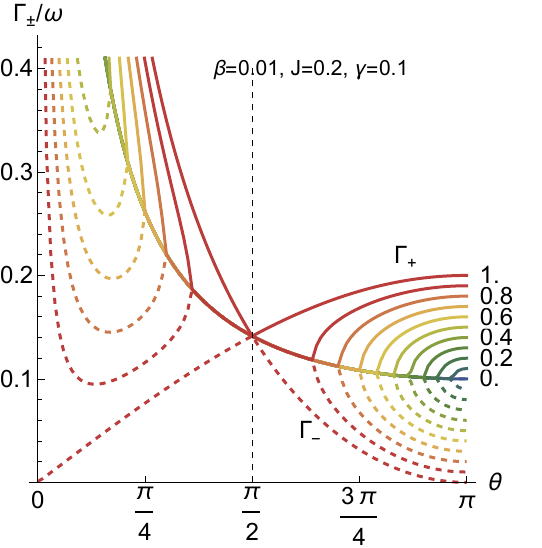}}
    \caption{
Mode relaxation rates $\Gamma_+$ and $\Gamma_-$ for a pair of anyonic oscillators as a function of the statistical angle $\theta$, shown at two reduced temperatures. (a) Low-temperature regime ($\beta \hbar \omega > 1$); (b) High-temperature regime ($\beta \hbar \omega \ll 1$). Solid and dashed curves indicate symmetric and antisymmetric mode decay rates, respectively. For $\xi < 0$, curve assignments are reversed. Parameters: $J = 0.2\omega$, $\gamma/\omega = 0.1$.
}
    \label{fig:2}
\end{figure*}

\subsection{Anyon double oscillator} 
To further generalize the analysis, consider the case of two anyon modes \( a_1 \) and \( a_2 \), 
coupled by an exchange term $J$, with Hamiltonian
\begin{align}
H/\hbar &= \omega ( a_{1}^{\dagger} a_{1} + a_{2}^{\dagger}a_{2}) + J(a_{1}^{\dagger}a_{2} + a_{2}^{\dagger}a_{1}).
\end{align}
We can bring $H$ into diagonal form by 
defining new anyon operators:
\(
\tilde b_\pm= (a_1 \pm e^{\pm i\theta} a_2)/\sqrt{2}
\)
such that their commutation relation 
preserves 
\begin{align}
    [\tilde b_\pm,\tilde b_\pm^\dagger]
    \approx (e^{i\theta N_1} + e^{i\theta N_2})/2
\end{align}
reflecting the non-local averaging over the anyonic occupation numbers.  
These bring the Hamiltonian into diagonal form
\begin{equation}
\tilde H/\hbar =\sum_{s=\pm} \Omega_s \tilde b_s^\dagger \tilde b_s
\end{equation}
with normal mode frequencies 
$
\omega_\pm = \omega \pm J \cos(\theta)
$.
Following along the lines of our previous works\cite{Bittner2024b,Tyagi:2024aa,Bittner:2025aa}
 we define Lindblad jump operators in terms of the deformed normal mode operators as  
\begin{align}
L_k = \lambda_k^{(+)} \tilde b_+
    + \lambda_k^{(-)} \tilde b_-
\end{align} with explicit coefficients
$\lambda_k^{(\pm)}$ for the emission 
and absorption channels are  given
in the Appendix \ref{Appendix:B}.
Using these, we derive the equations of motion 
for the deformed anyonic operators which we write in matrix form as
\begin{equation}
\frac{d}{dt}
\begin{pmatrix} \langle \tilde{b}_+ \rangle \\ \langle \tilde{b}_- \rangle \end{pmatrix} = W_{\mathrm{eff}} \cdot \begin{pmatrix} \langle \tilde{b}_+ \rangle \\ \langle \tilde{b}_- \rangle \end{pmatrix},
\end{equation}
where the effective evolution matrix
is given by
\begin{equation}
W_{\text{eff}} =
\begin{pmatrix}
-i \omega_+ - \Gamma_{++} & -\Gamma_{+-} \\
-\Gamma_{-+} & -i \omega_- - \Gamma_{--}
\end{pmatrix}.
\end{equation}
The $\Gamma_{ss'}$ coefficients determine the rate contributions to the adjoint dissipator matrix \( W_{\text{eff}} \), governing the mode-mode coherence decay. The statistical phase \( \theta \) enters through the relative phase factor \( e^{-i\theta/2} \), and \( \xi \) modulates the strength of symmetric versus antisymmetric 
environmental coupling and hence
determine whether or not a 
particular mode is enhanced or supressed depending upon the statistical phase and the environmental correlation.  Diagonalizing $W_{\text{eff}}$ gives the oscillation frequency and 
decay rates for the coherent evolution
of each mode. 
\begin{figure*}
    \centering
       \subfigure[]{\includegraphics[width=0.48\linewidth]{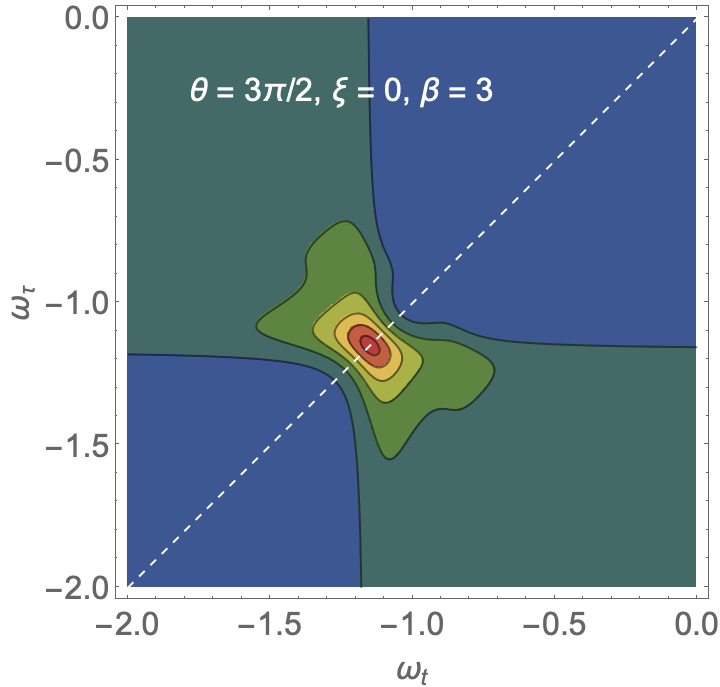}}
    \subfigure[]{\includegraphics[width=0.48\linewidth]{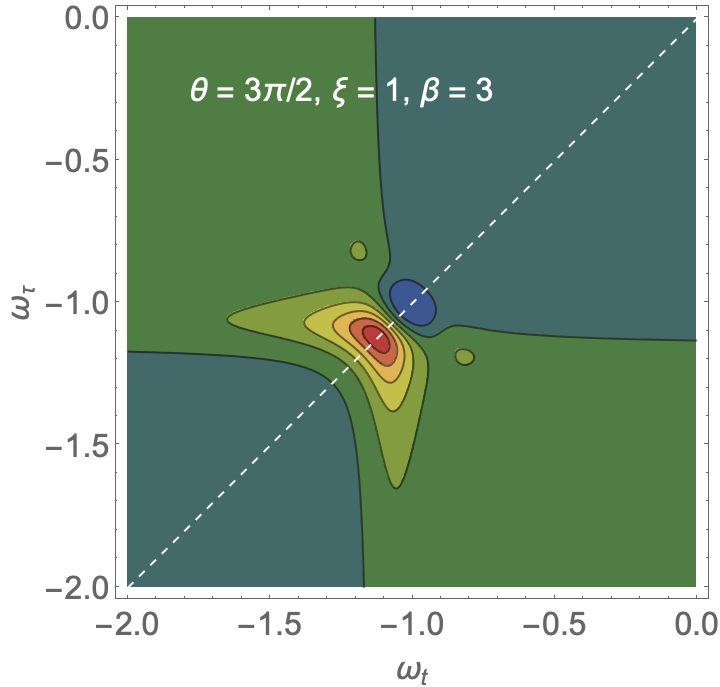}}
 
    \subfigure[]{\includegraphics[width=0.48\linewidth]{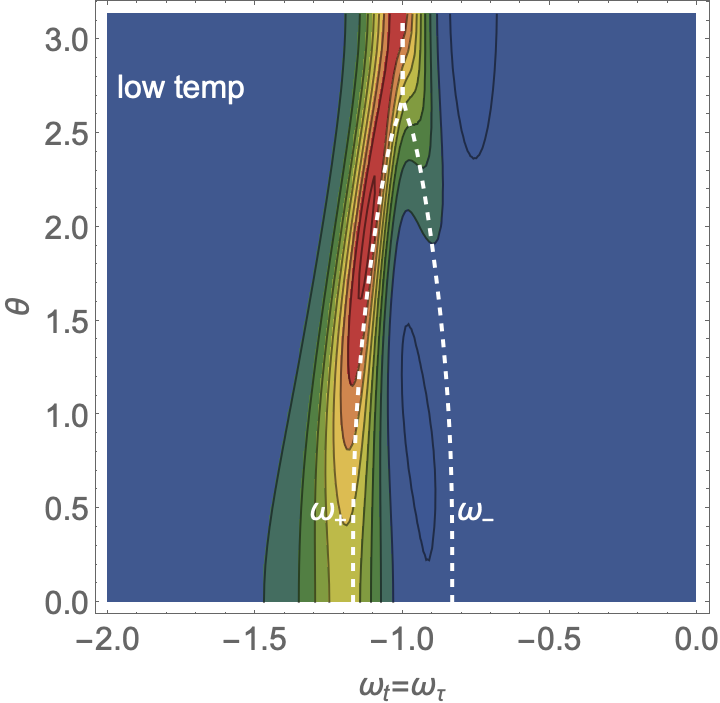}}
    \subfigure[]{\includegraphics[width=0.48\linewidth]{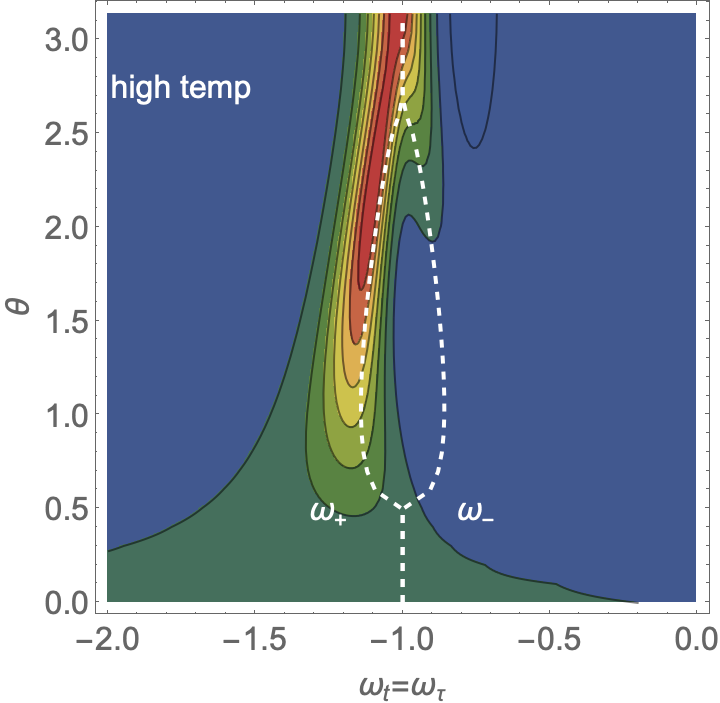}}
    \caption{
Predicted two-dimensional rephasing spectra from our theoretical model of an anyonic dimer system. (a) Real part of the third-order response in the normal regime ($\xi = 0$); (b) Symmetry-broken regime ($\xi = 1$), both computed at low temperature. (c, d) Evolution of the $\omega_t = \omega_\tau$ lineshape as the statistical angle $\theta$ varies from $0$ to $\pi$. Dashed overlays track the real parts of the normal mode frequencies, illustrating bifurcation across exceptional points. These theoretical results highlight spectral signatures that may be accessible via coherent nonlinear spectroscopy.
}
    \label{fig:3}
\end{figure*}

In Fig.~\ref{fig:2}(a,b) we show the 
total relaxation rates for the anyon dimer 
as we vary the statistical angle from 0 to $\pi$
and the correlation parameter, $-1\le \xi \le 1$. 
Parameters are chosen to correspond to physically reasonable
values of the hopping integral and relaxation rates. 
(Note for $\xi < 0$, dashed and solid curves are swapped.)
For $\theta < 2$, both normal modes of the systems 
relax at the same rate.  
However, for $\theta >2$ the 
two rates bifurcate and concurrently the two frequencies coalesce, indicating the presence of an exceptional point
that moves towards higher values of $\theta$ as the 
correlation parameter weakens.  At high temperature, 
we see both an upper and lower critical value of $\theta$ for a given $\xi$. On the bosonic side ($\theta < \pi/2$),
the relaxation rate is dominated by the purely thermodynamic contribution as per Eq.~\ref{eq:8}.  
As temperature increases, the lower point moves towards 
higher values of the statistical angle as $\xi$ increases.
On the fermionic side, $\theta > \pi/2$, the statistical 
contribution dominates and the upper critical value shifts
towards the left. At very high temperatures, the two
meet at $\theta = \pi/2$ suggesting that 
Majorana quasi-particles are insensitive to environmental correlation effects.

In the broken symmetry regime, one of the normal 
modes becomes more strongly damped while the other becomes
increasing more weakly damped as $\xi\to \pm 1$. 
In general for our correlated noise model ($\xi>0$) protects the 
anti-symmetric mode while the symmetric normal mode
is increasingly overdamped and vice-versa when $(\xi<0)$, 
as we discussed in our previous works\cite{Bittner:2025aa,Tyagi:2024aa,Bittner2024b}.
Hence, in the broken symmetry regime, the two particles 
become synchronized (or anti-synchronized). 

\subsection{Detection via non-linear spectroscopy}
Recent theoretical studies have established nonlinear spectroscopy as a powerful tool for detecting the signatures of anyonic statistics in topologically ordered systems. In particular, pump–probe protocols have been shown to reveal braiding-induced phase shifts in the dynamical response functions of two-dimensional systems hosting anyons \cite{McGinley2024PRL, Yang2025}. These braiding effects introduce universal features in the nonlinear signal that persists under realistic conditions, including finite temperature and non-universal interactions. In the toric code, for example, the nonlinear response distinguishes trivial from non-trivial anyon braiding, while in broader classes of topological phases, anomalous thermal relaxation and robust phase coherence provide experimentally accessible indicators of fractional statistics \cite{McGinley2024Arxiv}. These findings support the feasibility of using multidimensional coherent spectroscopies to probe and control statistical phases, aligning with our proposal to resolve mode-selective decoherence pathways in anyonic oscillator networks.

Coherent multidimensional spectroscopy offers a powerful means to resolve the correlations and dynamics of anyonic systems, where exchange statistics deviate from standard bosonic or fermionic behavior. Unlike linear techniques, which conflate homogeneous and inhomogeneous broadening, coherent probes using phase-locked pulse sequences can isolate intrinsic dephasing and reveal pathways 
that may be sensitive to fractional statistics.\cite{Kandada2019,Kandada2020,Kandada2022}

In a typical three-pulse 2D spectroscopy experiment, the system is driven by a sequence of phase-stabilized laser pulses with well-defined wave vectors $\mathbf{k}_1$, $\mathbf{k}_2$, and $\mathbf{k}_3$. The emitted nonlinear signal appears in specific directions determined by the phase-matching condition, which enforces conservation of photon momentum among the optical fields. For the rephasing signal--often associated with a photon echo--the signal wave vector is given by $\mathbf{k}_{\text{sig}} = -\mathbf{k}_1 + \mathbf{k}_2 + \mathbf{k}_3$. This particular combination isolates Liouville pathways that involve evolution with negative time-ordering in the first interval, enabling partial cancellation of inhomogeneous broadening. It is important to emphasize that these wave vectors refer to the directions of the applied laser fields and are not related to any Bloch wave vectors or crystal momentum within the sample. The phase-matching geometry provides a versatile means of isolating specific dynamical pathways in the nonlinear response, independent of the underlying translational symmetry. For a comprehensive discussion of these techniques and phase-matching rules, we refer the reader to the foundational texts by Mukamel~\cite{Mukamel1995} and Hamm and Zanni~\cite{HammZanni2011}.

\begin{widetext}
    Focusing our attention on the rephasing third-order response, we write
\begin{align}
\label{Eq-15}
  \tilde{R}^{(3)}(\omega_3, t_2, \omega_1) = \left(\frac{i}{\hbar}\right)^3 \, \text{Tr} \left\{ \mu^{(-)} \tilde{\mathcal{G}}(\omega_\tau) \mu^{(+)} \mathcal{G}(t_2) \mu^{(+)} \tilde{\mathcal{G}}(\omega_t) \mu^{(-)} \rho_{\text{eq}} \right\}  
\end{align}
where
$\tilde{\mathcal{G}}(\omega)$ is the resolvent of the Liouvillian, with $\omega_t$ and $\omega_\tau$ being the Fourier-conjugate variables of $t_3$ and $t_1$ respectively, and $\mathcal{G}(t_2) = e^{-i \mathcal{L} t_2}$ remains in the time domain and 
$\mu^{(\pm)}$ are superoperators representing the 
light-matter interaction. The $\pm$ indicates whether they
act to the left or right and $\rho_{\text{eq}}$ is the 
equilibrium density matrix.\cite{Mukamel1995}  However, we have to point out that the dipole interaction $\mu^{(\pm)}$ also carries 
information concerning the statistical phase. This is because in a two-site system, where $a_1^\dagger$, $a_2^\dagger$ are creation operators for modes 1 and 2, the anyon dipole operator can be written as
\[
\mu \propto  a_1^\dagger + a_1 +  e^{i \theta \hat{n}_1} a_2^\dagger +  e^{-i \theta \hat{n}_1} a_2
\]
whereby $e^{\pm i \theta \hat{n}_1}$ implements the braiding/statistical phase acquired by an anyon on site 2 when commuting past site 1.  Consequently, we anticipate 
signatures of this to be present in the 3rd-order rephasing signals. 

To interpret the physical content of Eq.~(\ref{Eq-15}), we recall that the rephasing pathway corresponds to a specific time-ordered sequence of interactions between the light field and the system. Starting from the thermal equilibrium density matrix $\rho_{\mathrm{eq}} = |0\rangle\langle 0|$, the system evolves through a sequence of coherence and population states induced by successive light–matter interactions. In superoperator notation, which tracks the action on the ket and bra sides of the density matrix, the rephasing pathway can be expressed as:
\begin{equation}
\rho^{(0)} 
\xrightarrow{\mu^{(-)}} 
|1\rangle\langle 0| 
\xrightarrow{\mathcal{G}(t_1)} 
\xrightarrow{\mu^{(+)}} 
|1\rangle\langle 1| 
\xrightarrow{\mathcal{G}(t_2)} 
\xrightarrow{\mu^{(+)}} 
|2\rangle\langle 1| 
\xrightarrow{\mathcal{G}(t_3)} 
\xrightarrow{\mu^{(-)}} 
|2\rangle\langle 0|
\end{equation}
Here, $\mu^{(+)}$ and $\mu^{(-)}$ are superoperators acting on the ket and bra sides, respectively, and $\mathcal{G}(t)$ denotes the time-evolution superoperator that includes both coherent Hamiltonian evolution and dissipative dynamics. The kets $|1\rangle$ and $|2\rangle$ represent excited states generated by the action of the $\mu^{(\pm)}$ operators and are generally superpositions of the symmetric and antisymmetric normal modes. The structure of these transitions and the resulting lineshape are directly influenced by the statistical phase factors that modify the matrix elements of $\mu$ in the anyonic basis. These phase-dependent modifications are expected to produce measurable signatures in the rephasing 2D spectra.

\end{widetext}
Fig.\ref{fig:3}(a,b) shows the 
real parts of the 3rd-order rephasing response
for a model set of parameters corresponding to the 
normal regime with $\xi= 0$ to the
the symmetry broken regime 
with $\xi = 1$.  Within the normal regime, the 
diagonal slice with $\omega_t = \omega_\tau$ has
the correct absorptive line-shape profile expected
from the Kramers-Kronig relations.  However, 
in the broken-symmetry regime the profile is 
far more asymmetric and takes a more ``derivative'' or 
dispersive profiles.  This profile 
is strongly dependent upon the statistical angle
as seen in Fig.~\ref{fig:3}(c,d) where 
we examine the line-shape profile from the bosonic to 
fermionic limits at both low and high temperatures.
Here, we note that not only the line-shape profile changes
from absorptive to dispersive, the peak position 
tracks the eigenvalue of the ``bright'' (dipole allowed) mode.  
One can also detect evidence of the mode bifurcation as the system passes through the various exceptional points. 
We emphasize that while the symmetry breaking 
is dependent upon the statistical phase, its origins
are  due to the correlated environment terms 
which introduce mixing between the symmetric $b_+$ and 
anti-symmetric $b_-$ modes of the system via $W_{\text{eff}}$. 
The primary effect of the fractional statistics 
is to enhance this mixing via the off-diagonal terms 
in the dynamical matrix.

\section{Discussion}

The integration of anyonic fractional statistics results in alterations to the dissipation framework, thereby influencing the stability of protected modes. Unlike bosonic systems, which facilitate exact mode protection, anyonic correlations introduce additional coupling terms that modify the dynamics. In the context of bosonic systems ($\theta = 0$), the criterion for protected modes is satisfied at $\xi = \pm 1$, ensuring that one mode is entirely decoupled from the environment. In contrast, for anyonic oscillators ($\theta \neq 0$), the dissipation structure is altered in such a manner that the protected mode is not entirely decoupled but acquires slight coupling to the environment due to cross terms induced by anyonic statistical phases. The dissipation does not discernibly decompose into symmetric and antisymmetric modes, resulting in partial leakage of the protected mode. As a result, exact mode protection is undermined, except in the bosonic ($\theta = 0$) or fermionic ($\theta = \pi$) limits.

Fractional statistics deform the dissipative dynamics of coupled oscillators, allowing tunable mode protection and coherence control. Unlike bosonic or fermionic systems, anyonic oscillators have phase-sensitive decoherence rates and exceptional points that scale with temperature. This suggests a strategy to use exchange statistics and environmental correlations to create dissipation-resilient states in topological quantum devices.

 Our findings demonstrate that the exchange phase of anyons can be leveraged as a tunable parameter for modulating decoherence pathways, suggesting a new approach to coherence control via statistical engineering in open topological platforms.
 Further, we argue that coherent spectroscopic probes 
offer a wealth of detailed insight into the subtle interplay between the statistical phase and correlations
within the environment.

The bifurcation between symmetric and antisymmetric mode lifetimes observed in our model reflects a dynamical symmetry-breaking process analogous to spontaneous $\mathcal{PT}$-symmetry breaking in non-Hermitian quantum systems, as the eigenvalues of the dynamical matrix are no longer complex conjugate symmteric after symmetry breaking. When the environmental coupling becomes strongly correlated ($\xi \to \pm 1$), the Liouvillian spectrum exhibits exceptional points—non-Hermitian degeneracies at which both eigenvalues and eigenvectors coalesce—leading to the emergence of dominant, long-lived modes. This phenomenon manifests as synchronization or antisynchronization of the coupled oscillators, depending on the sign of the bath correlation. The anyonic exchange phase enhances this symmetry breaking by modulating the off-diagonal terms in the effective dynamical matrix $W_{\mathrm{eff}}$, thereby shifting the location and character of the exceptional points. These findings connect naturally to broader efforts to characterize noise-induced phase locking in open quantum systems~\cite{Bittner:2025aa,Tyagi:2024aa,Huber2020PT}, and they echo the non-Hermitian transitions studied in optical and condensed matter platforms~\cite{Ashida2020}. We believe that fractional statistics provide a novel and tunable axis for controlling dissipative phase transitions and coherence lifetimes in topological matter.

\begin{acknowledgments}
This work was supported by the National Science Foundation (CHE-2404788) and the Robert A. Welch Foundation (E-1337). The authors thank 
Andrei Piryatinski (LANL), 
Ajay Kandada (Wake Forest), and Carlos Silva (U Montreal) for useful discussions pertaining to the proposed non-linear optical experiments.
\end{acknowledgments}

\section*{Data Availability Statement}
Appendix includes full derivations of the Lindblad dynamics, analytical forms of the statistical decay rate, and a Mathematica notebook reproducing all plots and simulations.
Mathematica notebooks 
for the figures are available online at
\href{https://www.wolframcloud.com/obj/01f1cb8b-2d95-4b37-99c6-a0f366a91d52}{this Wolfram Cloud link}.
An accompanying WolframLanguage 
package is available 
\href{https://www.wolframcloud.com/obj/2de91f32-122d-4fd8-b2be-2598401ea88e}{at this link}.

\section*{Competing Interests}

The authors declare no competing interests.

\section*{Author Contributions}
E.R.B. conceived the project and developed the theoretical framework. B.T. helped with the algebraic derivations and numerical simulations. Both authors contributed to the interpretation of results and writing of the manuscript.

\bibliographystyle{apsrev4-2}  
\bibliography{References_local}

\newpage
\appendix
\section{Single oscillator dynamics\label{Appendix:A}
}

This section provides detailed derivations of key expressions appearing in the main text, including the statistical contribution to the relaxation rate, the adjoint form of the Lindblad dissipator for a single anyonic oscillator, and the functional form of the exchange-phase expectation value $\langle e^{i\theta N} \rangle$. These results underlie the coherence dynamics of the system and clarify how the anyonic exchange statistics modify both thermal occupation and dissipation. While the main text presents compact formulas suitable for presentation, we include these step-by-step derivations here to assist readers seeking deeper insight into the underlying calculations. In particular, we aim to support reviewers and readers who may be unfamiliar with statistical-phase averaging or with how Lindblad dynamics are extended to systems obeying deformed commutation relations.
\subsection{Anyon Commutation Structure}

We first recall the definition of the deformed algebra:
\begin{align*}
    aa^\dagger - e^{i\theta}a^\dagger a &= 1
\end{align*}
So the commutator,
\begin{align*}
    [a, a^\dagger] &= a a^\dagger - a^\dagger a = 1 + e^{i \theta} N-N,\\
    [a, a^\dagger] &= 1 +(e^{i \theta}-1)N \equiv \Phi (\hat{N})
\end{align*}
where $\hat{N} = a^\dagger a.$ is the number operator.
In the bosonic limit $(\theta = 0)$:
\[
\Phi(\hat{N}) = 1.
\]
In the fermionic limit (\(\theta = \pi\)):
\[
\Phi(\hat{N}) = 1 - 2\hat N.
\]

\subsection{Statistical Phase Averaging}

To capture the effect of anyonic exchange statistics on coherence decay, we compute the thermal expectation of the phase factor
\[
\langle e^{i\theta \hat N} \rangle = \sum_{n=0}^\infty e^{i\theta n} \, P_n,
\]
where  $P_n = (1 - z) z^n$ is the canonical probability for occupying the n-th level in a harmonic mode with $z = e^{-\beta \hbar \omega}$. Recognizing the sum as a geometric series, we obtain the closed-form expression
\[
\langle e^{i\theta \hat N} \rangle = \frac{1 - z}{1 - z e^{i\theta}},
\]
whose real part becomes
\[
\text{Re}\langle e^{i\theta \hat N} \rangle = \frac{(1 - z)(1 - z \cos\theta)}{1 - 2z \cos\theta + z^2}.
\]
As we show below, the
statistical contribution to the relaxation rate then naturally follows as
\[\Gamma_{\text{stat}} = \frac{\gamma}{2} \left(1 - \text{Re}\langle e^{i\theta \hat N} \rangle\right),\]
highlighting how the thermal population distribution and anyonic phase combine to suppress coherence.

\subsection{Heisenberg equations of motion}
We begin with the diagonal Hamiltonian for a single anyonic mode,
\[
H = \hbar\omega\, \hat{N}, \quad \text{where } \hat{N} = a^\dagger a.
\]
The Heisenberg equation of motion for the annihilation operator \( a \) is given by
\[
\frac{d a}{dt}  = \frac{i}{\hbar} [H, a] = i\omega [\hat{N}, a].
\]
To evaluate this, we use the identity
\[
[\hat{N}, a] = a^\dagger [a, a] + [a^\dagger, a] a.
\]
Since \( [a, a] = 0 \) and \( [a^\dagger, a] = \Phi(\hat{N}) \), this reduces to
\[
[\hat{N}, a] = -\Phi(\hat{N}) a.
\]
Hence,
\[
\frac{d}{dt} a = -i\omega \Phi(\hat{N}) a.
\]
In general, the operator \( \Phi(\hat{N}) \) introduces a nonlinear correction to the coherent evolution due to the underlying exchange statistics. This term modifies the frequency of oscillation in a number-dependent way, encoding the statistical phase in the unitary dynamics.

However, for the purposes of deriving the relaxation rates, we approximate
\[
[H, a] \approx -\hbar\omega a,
\]
corresponding to the bosonic case where \( \Phi(\hat{N}) = 1 \). This approximation allows us to isolate the effects of dissipation and fractional statistics in the Lindblad term without introducing nonlinearities into the coherent evolution. The validity of this approximation is justified in the regime where the deformation from bosonic statistics is modest (i.e., \( \theta \ll \pi \)) or when the statistical effects enter dominantly through the dissipator. Numerical simulations confirm that this substitution introduces negligible error in the computed coherence decay rates for the physical parameter ranges considered.

\subsection{Dissipative terms}

To derive the total coherence relaxation rate for a single anyon oscillator, we begin with the Lindblad master equation for the density matrix \( \rho \), including both emission and absorption processes:
\begin{align}
\frac{d\rho}{dt} = -i[H, \rho] + L_{\text{em}} \rho L_{\text{em}}^\dagger - \frac{1}{2} \{L_{\text{em}}^\dagger L_{\text{em}}, \rho\} \nonumber \\+ L_{\text{abs}} \rho L_{\text{abs}}^\dagger - \frac{1}{2} \{L_{\text{abs}}^\dagger L_{\text{abs}}, \rho\}.
\end{align}
Here, the Hamiltonian is \( H = \omega a^\dagger a \), and the Lindblad operators are:
\begin{equation}
L_{\text{em}} = \sqrt{\gamma(n_\theta + 1)}\, a, \quad L_{\text{abs}} = \sqrt{\gamma n_\theta}\, a^\dagger.
\end{equation}

Moving to the Heisenberg picture, the evolution of an operator \( \hat{O} \) is governed by:
\begin{equation}
\frac{d\hat{O}}{dt} = i[H, \hat{O}] + \mathcal{D}^\dagger_{\text{em}}[\hat{O}] + \mathcal{D}^\dagger_{\text{abs}}[\hat{O}],
\end{equation}
where the adjoint dissipator is
\begin{equation}
\mathcal{D}^\dagger[L][\hat{O}] = \frac{1}{2} \left( L [\hat{O}, L^\dagger] + [L, \hat{O}] L^\dagger \right).
\end{equation}

We apply this to the annihilation operator \( a \). The Hamiltonian contribution is
\begin{equation}
i[H, a] = -i \omega a.
\end{equation}

The dissipator terms yield:
\begin{align}
\mathcal{D}^\dagger_{\text{em}}[a] &= \frac{1}{2} \gamma(n_\theta + 1) \Phi(\hat{N})a, \\
\mathcal{D}^\dagger_{\text{abs}}[a] &= -\frac{1}{2} \gamma n_\theta \Phi(\hat{N})a,
\end{align}
so that the total dissipative term becomes:
\begin{equation}
\mathcal{D}^\dagger[a] = \mathcal{D}_{\text{em}}^\dagger + \mathcal{D}_{\text{abs}}^\dagger = \frac{1}{2} \gamma \Phi(\hat{N})a
\end{equation}
We approximate the operator-valued $\Phi(\hat{N})$ by its thermal average, allowing the statistical effects to be encoded in a scalar factor.
\begin{equation}
\frac{d a}{dt} = -i \omega a + \frac{\gamma}{2} \langle e^{i N \theta} \rangle a.
\end{equation}
Here, we assume the unitary contribution remains linear in $a$ as discussed in Section 1.3, and we retain the full statistical correction via the dissipator.

Taking the expectation value over a thermal ensemble and including both thermal noise and statistical interference, the full coherence equation becomes:
\begin{equation}
\frac{d}{dt} \langle a \rangle = \left( -i \omega - \Gamma_\text{full} \right) \langle a \rangle,
\end{equation}
where the total phase relaxation rate is
\begin{equation}
\Gamma_\text{full} = \frac{\gamma}{2} \left[ 2 n_\theta + 1 + \left( 1 - \text{Re} \langle e^{i N \theta} \rangle \right) \right].
\end{equation}

\textbf{Remark.}
The adjoint form of the Lindblad dissipator, $\mathcal{D}^\dagger[A]$, is valid strictly under the trace operation, i.e., when computing expectation values of observables via $\langle A \rangle = \mathrm{Tr}[\rho\, A]$. It is derived to satisfy the dual relation $\mathrm{Tr}[\rho\, \mathcal{D}^\dagger[A]] = \mathrm{Tr}[(\mathcal{D}[\rho])\, A]$. Outside of the trace, $\mathcal{D}^\dagger[A]$ may not preserve the algebraic structure of $A$, especially in systems with deformed commutation relations. For this reason, care must be taken when interpreting $\mathcal{D}^\dagger$ as an operator-valued object rather than a generator of expectation value dynamics.

\subsubsection{Limiting Cases: Bosons and Fermions}

\textbf{Bosonic limit ($\theta = 0$).}
In the limit $\theta \to 0$, the anyonic algebra reduces to the canonical bosonic commutation relation $[a, a^\dagger] = 1$. The statistical average becomes:
\[
\langle e^{i\theta N} \rangle \to 1, \quad \text{so} \quad \Gamma_{\text{stat}} \to 0,
\]
and the full relaxation rate reduces to the thermal contribution alone:
\[
\Gamma_{\text{full}}^{\text{(boson)}} = \frac{\gamma}{2}(2n + 1),
\]
where $n = \langle N \rangle = \frac{1}{e^{\beta \hbar \omega} - 1}$. This result matches the textbook expression for the decay rate of $\langle a \rangle$ in a damped bosonic mode coupled to a thermal bath (see Breuer \& Petruccione, Carmichael, Gardiner \& Zoller).

\textbf{Fermionic limit ($\theta = \pi$).}
In the fermionic limit, the algebra becomes $[a, a^\dagger] = 1 - 2\hat{N}$ and the occupation number is $n_F = \langle N \rangle = \frac{1}{e^{\beta \hbar \omega} + 1}$. The statistical phase average is:
\[
\langle e^{i\pi N} \rangle = \sum_n (-1)^n P_n = 1 - 2n_F,
\]
so the statistical contribution becomes:
\[
\Gamma_{\text{stat}} = \frac{\gamma}{2}\left(1 - (1 - 2n_F)\right) = \gamma n_F.
\]
The thermal part becomes $\frac{\gamma}{2}(2n_F + 1)$, yielding:
\[
\Gamma_{\text{full}}^{\text{(fermion)}} = \frac{\gamma}{2}(2n_F + 1) + \gamma n_F = \frac{\gamma}{2}(2n_F + 1) + \gamma n_F.
\]
This simplification reflects the exclusion principle, where $n_F \leq 1$ and higher occupation is forbidden. The resulting decay rate is reduced compared to the bosonic case at low temperatures but saturates at high $T$.

\textbf{Summary.}
These limiting cases validate the general anyonic expression for $\Gamma_{\text{full}}$ and illustrate how the statistical contribution smoothly interpolates between enhanced decoherence (bosons) and suppressed coherence decay (fermions) depending on $\theta$ and $T$.
\section{Anyonic Algebra for the 2-oscillator system\label{Appendix:B}}
To accurately account for the anyonic algebra in the two-oscillator system, we define deformed normal mode operators that incorporate the statistical phase \( \theta \). Starting from the local anyon operators \( a_1 \), \( a_2 \) satisfying
\begin{equation}
a_i a_j^\dagger - e^{-i \theta} a_j^\dagger a_i = \delta_{ij},
\end{equation}
we define symmetric and antisymmetric deformed normal modes:
\begin{equation}
\tilde{b}_\pm = \frac{1}{\sqrt{2}} \left( a_1 \pm e^{i\theta/2} a_2 \right), \quad
\tilde{b}_\pm^\dagger = \frac{1}{\sqrt{2}} \left( a_1^\dagger \pm e^{-i\theta/2} a_2^\dagger \right).
\end{equation}

The commutation relations for these operators become
\begin{equation}
[\tilde{b}_\pm, \tilde{b}_\pm^\dagger] \approx \frac{1}{2} \left( e^{i\theta N_1} + e^{i\theta N_2} \right),
\end{equation}
reflecting the nonlocal averaging over the anyonic occupation numbers. This structure preserves the fractional statistics while defining effective mode operators.

We now re-express the Hamiltonian:
\begin{equation}
H = \omega (a_1^\dagger a_1 + a_2^\dagger a_2) + J (a_1^\dagger a_2 + a_2^\dagger a_1)
\end{equation}
in terms of \( \tilde{b}_\pm \). Substituting the inverse relations and simplifying, we obtain
\begin{equation}
H = \omega_+ \tilde{b}_+^\dagger \tilde{b}_+ + \omega_- \tilde{b}_-^\dagger \tilde{b}_-,
\end{equation}
with effective normal mode frequencies:
\begin{equation}
\omega_\pm = \omega \pm J \cos(\theta/2).
\end{equation}
Thus, the statistical phase \( \theta \) modulates the mode splitting through a cosine factor, recovering the bosonic result in the limit \( \theta \to 0 \).

Next, we re-express the Lindblad operators originally defined as correlated combinations of \( a_1 \) and \( a_2 \):
\begin{equation}
L^\pm_{\text{em}} = \sqrt{\gamma(n_\theta + 1)} \sqrt{1 \pm \xi} \cdot \frac{a_1 \pm a_2}{\sqrt{2}},
\end{equation}
in terms of the deformed basis:
\begin{equation}
a_1 \pm a_2 = \frac{1}{\sqrt{2}} (\tilde{b}_+ \pm e^{-i\theta/2} \tilde{b}_-),
\end{equation}
leading to the transformed Lindblad operators:
\begin{equation}
L^\pm_{\text{em}} = \sqrt{\gamma(n_\theta + 1)} \sqrt{1 \pm \xi} \cdot \frac{1}{2} \left( \tilde{b}_+ \pm e^{-i\theta/2} \tilde{b}_- \right),
\end{equation}
and similarly for the absorption operator:
\begin{equation}
L^\pm_{\text{abs}} = \sqrt{\gamma n_\theta} \sqrt{1 \pm \xi} \cdot \frac{1}{2} \left( \tilde{b}_+ \pm e^{-i\theta/2} \tilde{b}_- \right).
\end{equation}

These expressions preserve the algebraic structure of the anyonic system while allowing coherent dynamics to be described in the diagonal deformed basis. The phase factors embedded in the transformation ensure that the dissipative terms remain consistent with the underlying statistics.

To determine the equations of motion in the deformed basis, we begin with the Heisenberg equation for an operator \( \hat{O} \):
\begin{equation}
\frac{d\hat{O}}{dt} = i[H, \hat{O}] + \sum_k \mathcal{D}^\dagger_k[\hat{O}],
\end{equation}
where \( H = \omega_+ \tilde{b}_+^\dagger \tilde{b}_+ + \omega_- \tilde{b}_-^\dagger \tilde{b}_- \), with \( \omega_\pm = \omega \pm J \cos(\theta/2) \). The Lindblad operators in the deformed basis are:
\begin{align}
L^\pm_{\text{em}} &= \sqrt{\gamma(n_\theta + 1)} \sqrt{1 \pm \xi} \cdot \frac{1}{2} \left( \tilde{b}_+ \pm e^{-i\theta/2} \tilde{b}_- \right), \\
L^\pm_{\text{abs}} &= \sqrt{\gamma n_\theta} \sqrt{1 \pm \xi} \cdot \frac{1}{2} \left( \tilde{b}_+ \pm e^{-i\theta/2} \tilde{b}_- \right).
\end{align}

Evaluating the commutator terms gives the coherent evolution:
\begin{equation}
i[H, \tilde{b}_\pm] = -i \omega_\pm \tilde{b}_\pm.
\end{equation}

The dissipative terms contribute both diagonal and off-diagonal damping:
\begin{align}
\frac{d \tilde{b}_+}{dt} &= -i \omega_+ \tilde{b}_+ - \Gamma_{++} \tilde{b}_+ - \Gamma_{+-} \tilde{b}_-, \\
\frac{d \tilde{b}_-}{dt} &= -i \omega_- \tilde{b}_- - \Gamma_{--} \tilde{b}_- - \Gamma_{-+} \tilde{b}_+,
\end{align}
where the coefficients \( \Gamma_{ij} \) arise from the overlap structure of the Lindblad operators:
\begin{equation}
\Gamma_{ij} = \sum_k \frac{1}{2} \left( |\lambda_k^{(i)}|^2 + |\lambda_k^{(j)}|^2 \right).
\end{equation}
To evaluate the dissipative coupling terms \( \Gamma_{ij} \), we express each Lindblad operator in the deformed basis as a linear combination of \( \tilde{b}_+ \) and \( \tilde{b}_- \):
\begin{equation}
L_k = \lambda_k^{(+)} \tilde{b}_+ + \lambda_k^{(-)} \tilde{b}_-.
\end{equation}

The explicit coefficients \( \lambda_k^{(i)} \) for each emission and absorption channel are given as follows. First for the {Emission channels:}
\begin{align}
\lambda_+^{(+)} &= \sqrt{\gamma(n_\theta + 1)} \cdot \frac{\sqrt{1 + \xi}}{2}, \label{Eq:B19}\\
\lambda_+^{(-)} &= \sqrt{\gamma(n_\theta + 1)} \cdot \frac{\sqrt{1 + \xi}}{2} \cdot e^{-i\theta/2}, \\
\lambda_-^{(+)} &= \sqrt{\gamma(n_\theta + 1)} \cdot \frac{\sqrt{1 - \xi}}{2}, \\
\lambda_-^{(-)} &= -\sqrt{\gamma(n_\theta + 1)} \cdot \frac{\sqrt{1 - \xi}}{2} \cdot e^{-i\theta/2}.
\end{align}
For the \textbf{Absorption channels:}
\begin{align}
\lambda_+^{(+)} &= \sqrt{\gamma n_\theta} \cdot \frac{\sqrt{1 + \xi}}{2}, \\
\lambda_+^{(-)} &= \sqrt{\gamma n_\theta} \cdot \frac{\sqrt{1 + \xi}}{2} \cdot e^{-i\theta/2}, \\
\lambda_-^{(+)} &= \sqrt{\gamma n_\theta} \cdot \frac{\sqrt{1 - \xi}}{2}, \\
\lambda_-^{(-)} &= -\sqrt{\gamma n_\theta} \cdot \frac{\sqrt{1 - \xi}}{2} \cdot e^{-i\theta/2}.
\label{Eq:B26}
\end{align}
These coefficients determine the rate contributions to the adjoint dissipator matrix \( W_{\text{deformed}} \), governing the mode-mode coherence decay. The statistical phase \( \theta \) enters through the relative phase factor \( e^{-i\theta/2} \), and \( \xi \) modulates the strength of symmetric versus antisymmetric environmental coupling.

Defining the operator vector \( \vec{b}(t) = \begin{pmatrix} \tilde{b}_+ \\ \tilde{b}_- \end{pmatrix} \), the equations of motion assume the matrix form:
\begin{equation}
\frac{d\vec{b}}{dt} = W_{\text{deformed}} \vec{b},
\end{equation}
with
\begin{equation}
W_{\text{deformed}} =
\begin{pmatrix}
-i \omega_+ - \Gamma_{++} & -\Gamma_{+-} \\
-\Gamma_{-+} & -i \omega_- - \Gamma_{--}
\end{pmatrix}.
\end{equation}

The statistical phase \( \theta \) and bath correlation \( \xi \) enter both the diagonal frequencies and the off-diagonal dissipative couplings, encoding interference between the deformed normal modes.

We now derive the equations of motion for the expectation values \( \langle \tilde{b}_\pm \rangle \) using the Heisenberg picture with the adjoint Lindblad dissipator. For each operator \( \tilde{b}_\pm \), the time evolution is given by
\begin{equation}
\frac{d}{dt} \tilde{b}_\pm = -i \omega_\pm \tilde{b}_\pm + \sum_k \mathcal{D}_k^\dagger[\tilde{b}_\pm],
\end{equation}
where the sum includes both emission and absorption channels indexed by \( k = +, - \).

Each Lindblad operator takes the form
\begin{equation}
L_k = \lambda_k^{(+)} \tilde{b}_+ + \lambda_k^{(-)} \tilde{b}_-,
\end{equation}
leading to the adjoint dissipator
\begin{equation}
\mathcal{D}_k^\dagger[\tilde{b}_+ ] = -\left( |\lambda_k^{(+)}|^2 \tilde{b}_+ + \lambda_k^{(+)} \lambda_k^{(-)*} \tilde{b}_- \right),
\end{equation}
and similarly for \( \tilde{b}_- \). Thus, the equations of motion for the first moments become
\begin{align}
\frac{d}{dt} \langle \tilde{b}_+ \rangle &= \left( -i \omega_+ - \sum_k |\lambda_k^{(+)}|^2 \right) \langle \tilde{b}_+ \rangle - \sum_k \lambda_k^{(+)} \lambda_k^{(-)*} \langle \tilde{b}_- \rangle, \\
\frac{d}{dt} \langle \tilde{b}_- \rangle &= \left( -i \omega_- - \sum_k |\lambda_k^{(-)}|^2 \right) \langle \tilde{b}_- \rangle - \sum_k \lambda_k^{(-)} \lambda_k^{(+)*} \langle \tilde{b}_+ \rangle.
\label{Eq:B35}
\end{align}

These coupled equations can be written in matrix form as
\begin{equation}
\frac{d}{dt}
\begin{pmatrix} \langle \tilde{b}_+ \rangle \\ \langle \tilde{b}_- \rangle \end{pmatrix} = W_{\text{eff}} \cdot \begin{pmatrix} \langle \tilde{b}_+ \rangle \\ \langle \tilde{b}_- \rangle \end{pmatrix},
\end{equation}
where the effective evolution matrix is
\begin{equation}
W_{\text{eff}} =
\begin{pmatrix}
-i \omega_+ - \sum_k |\lambda_k^{(+)}|^2 & -\sum_k \lambda_k^{(+)} \lambda_k^{(-)*} \\
-\sum_k \lambda_k^{(-)} \lambda_k^{(+)*} & -i \omega_- - \sum_k |\lambda_k^{(-)}|^2
\end{pmatrix}.
\end{equation}

This expression compactly encodes both the coherent dynamics and the dissipative coupling between the deformed normal modes arising from environmental correlations and anyonic statistics.
The coherence lifetimes \( \tau_\pm \) of the deformed normal modes \( \tilde{b}_\pm \) are defined by the inverse of the real parts of the eigenvalues \( \lambda_\pm \) of the effective matrix \( W_{\text{eff}} \):
\begin{equation}
\tau_\pm = \frac{1}{-\mathrm{Re}(\lambda_\pm)}.
\end{equation}
These quantities characterize the exponential decay of the mode coherences \( \langle \tilde{b}_\pm(t) \rangle \sim e^{\lambda_\pm t} \). When the off-diagonal couplings vanish, the eigenvalues reduce to the diagonal entries:
\begin{equation}
\lambda_+ = -i \omega_+ - \Gamma_{++}, \qquad \lambda_- = -i \omega_- - \Gamma_{--},
\end{equation}
and the corresponding lifetimes simplify to
\begin{equation}
\tau_+ = \frac{1}{\Gamma_{++}}, \qquad \tau_- = \frac{1}{\Gamma_{--}}.
\end{equation}
More generally, the full eigenvalues \( \lambda_\pm \) incorporate both decay and mode mixing contributions:
\begin{equation}
\lambda_\pm = \frac{1}{2} \left( A + D \pm \sqrt{(A - D)^2 + 4BC} \right),
\end{equation}
where the effective matrix components are
\begin{align}
A &= -i \omega_+ - \Gamma_{++}, & B &= -\Gamma_{+-}, \\
C &= -\Gamma_{-+}, & D &= -i \omega_- - \Gamma_{--}.
\end{align}
These eigenvalues determine the collective mode dynamics and coherence properties, with \( \tau_\pm \) describing the lifetime of each hybridized mode under the influence of statistical phase \( \theta \), bath asymmetry \( \xi \), and temperature-dependent occupation \( n_\theta \).
\section{Computational Notebook}

The accompanying Mathematica notebook linked in the 
Data Availability Statement contains the results of our symbolic derivations and numerical simulations corresponding to the data presented in Figures 1--3 of the main manuscript. The file includes parameter sweeps over the statistical angle $\theta$, dissipation rate calculations, and spectral response simulations based on the third-order susceptibility.
\textit{Mathematica} notebooks 
for the figures are available online at
\href{https://www.wolframcloud.com/obj/01f1cb8b-2d95-4b37-99c6-a0f366a91d52}{this Wolfram Cloud link}.
An accompanying WolframLanguage 
package is available as
\href{https://www.wolframcloud.com/obj/2de91f32-122d-4fd8-b2be-2598401ea88e}{here}.
\end{document}